\documentclass[twoside]{article}
\usepackage{fleqn,espcrc2}
\usepackage{amsmath}

\usepackage[dvips]{graphicx}

\newcommand{\etal}{{\it et al.}}

\newcommand{\be}{\begin{eqnarray}}
\newcommand{\ee}{\end{eqnarray}}
\newcommand{\bdm}{\begin{displaymath}}
\newcommand{\edm}{\end{displaymath}}

\def\Dslash{{D}\!\!\!\!/\,}

\def\spose#1{\hbox to 0pt{#1\hss}}
\def\ltapprox{\mathrel{\spose{\lower 3pt\hbox{$\mathchar"218$}}
 \raise 2.0pt\hbox{$\mathchar"13C$}}}
\def\gtapprox{\mathrel{\spose{\lower 3pt\hbox{$\mathchar"218$}}
 \raise 2.0pt\hbox{$\mathchar"13E$}}}
\def\inapprox{\mathrel{\spose{\lower 3pt\hbox{$\mathchar"218$}}
 \raise 2.0pt\hbox{$\mathchar"232$}}}

\title{High temperature QCD with three flavors of improved staggered quarks
\thanks{Presented by U.M.~Heller and R.L.~Sugar.}}
\author{ C.~Bernard
\address{Department of Physics, Washington University, St.~Louis, MO 63130,
USA},
T.~Burch
\address{Department of Physics, University of Arizona, Tucson, AZ 85721, USA}, 
C.E.~DeTar
\address{Physics Department, University of Utah, Salt Lake City, UT
  84112, USA},
Steven~Gottlieb
\address{Department of Physics, Indiana University, Bloomington, IN 47405,
USA},
Eric~Gregory$\,\null^{\rm b}$,
U.M.~Heller
\address{CSIT, Florida State University, Tallahassee, FL 32306-4120, USA},
J.~Osborn$\,\null^{\rm c}$,
R.L.~Sugar
\address{Department of Physics, University of California, Santa Barbara,
CA 93106, USA},
and D.~Toussaint$\,\null^{\rm b}$
} 

\begin{document}

\begin{abstract}
We present an update of our study of high temperature QCD with three
flavors of quarks, using a Symanzik improved gauge action and the Asqtad
staggered quark action. Simulations are being carried out on lattices
with $N_t=4$, 6 and 8 for the case of three degenerate quarks with masses
less than or equal to the strange quark mass, $m_s$, and on lattices with
$N_t=6$ and 8 for degenerate up and down quarks with masses in the range
$0.2 m_s \leq m_{u,d} \leq 0.6 m_s$, and the strange quark fixed near its
physical value. We also report on first computations of quark number
susceptibilities with the Asqtad action. These susceptibilities are of 
interest because they can be related to event-by-event
fluctuations in heavy ion collision experiments. 
Use of the improved quark action leads to a
substantial reduction in lattice artifacts. This can be seen already for
free fermions and carries over into our results for QCD.
\end{abstract}

\maketitle 

We are studying high temperature QCD with three flavors of
improved staggered quarks \cite{MILC_HT01} using a one--loop Symanzik
improved gauge action and the Asqtad quark action \cite{MILC_FATTEST,LEPAGE98}.
Both the gauge and quark actions have all lattice artifacts removed through
order $a^2$ ($a$ is the lattice spacing) at tree level, and are tadpole
improved. Thus, the leading order finite lattice spacing artifacts are of
order $\alpha_s a^2$, $a^4$. We consider two cases:
1) all three quarks have the same mass $m_q$; and 2) the two lightest
mass quarks have equal mass $m_{u,d}$ and the mass of the third
quark is fixed at that of the strange quark $m_s$. We refer to these cases
as $N_f=3$ and $N_f=2+1$, respectively. 

Because it contains the Naik term \cite{NAIK} the Asqtad action has a
significantly better dispersion relation for the quarks than the standard
Kogut--Susskind and Wilson quark actions, which helps decrease lattice
artifacts in the energy and pressure \cite{MILC_HT01}, and, as we shall
see, for the quark number susceptibilities. The Asqtad action also exhibits
excellent scaling properties \cite{IMP_SCALING}, and significantly better
flavor symmetry properties than the conventional Kogut--Susskind action
\cite{MILC_FATTEST,MILC_Spec}. We estimate that for lattices with eight to
ten time slices, the kaon will be heavier than the heaviest non-Goldstone
pion in the neighborhood of the finite temperature transition or crossover.

We have attempted to vary the temperature while keeping all other physical
quantities constant. To this end we have performed a set of spectrum
calculations at $a=0.13$~fm and 0.20~fm for three equal mass quarks.
Here, and throughout this work, we determine the lattice spacing from the
static $\bar Q Q$ potential, and express dimensionful quantities in terms of
$r_1$ defined by $r_1^2\,F_{\bar Q Q}(r_1)=1$, which leads to
$r_1\approx 0.34$~fm. We determine $m_s$ from the requirement that
$m_{\eta_{ss}}/m_\phi \approx 0.673$. We have performed spectrum calculations 
at both lattice spacings with $m_q=m_s$, $0.6\, m_s$ and $0.4\, m_s$,
with an additional calculation at $0.2\, m_s$ in progress. We would like to
carry out our thermodynamics studies with three equal mass quarks for
$m_{\eta_{ss}}/m_\phi$ fixed, but this quantity will, of course, vary
slightly with lattice spacing if we keep $m_q/m_s$ fixed. So, at $a=0.2$~fm
we perform linear interpolations of $m_{\eta_{ss}}^2$ and $m_\phi$ in the
quark mass to determine the precise values of $m_q$ for which
$m_{\eta_{ss}}/m_\phi$ takes on the values found at $a=0.13$~fm. Then
to determine the values of $am_q$ and $a$ for thermodynamics studies
with $0.13{\rm \ fm}< a < 0.20{\rm \ fm}$, we perform linear interpolations of
the logarithms of these quantities in the gauge coupling $10/g^2$.

Our approach for thermodynamics studies with $m_{u,d}< m_s$ is quite similar.
In this case we wish to vary the temperature keeping both $m_{\pi}/m_{\rho}$
and $m_{\eta_{ss}}/m_\phi$ fixed. We have therefore performed a set of spectrum
calculations at $a=0.13$~fm and 0.20~fm with the mass of the light quarks,
$m_{u,d}=0.6\, m_s$, $0.4\, m_s$ and $0.2\, m_s$. An additional spectrum
calculation with $m_{u,d}=0.1\, m_s$ is in progress. In these calculations
the mass of the heavy quark is fixed at $m_s$, as determined from spectrum
calculations with three equal mass quarks. In our spectrum runs at
$a=0.13$~fm \cite{MILC_Spec} we found that $m_{\eta_{ss}}$ varied by less than
2\% and $m_\phi$ by less than 1\% for $0.2\, m_s \leq m_{u,d}\leq m_s$
and the heavy quark mass held fixed at $m_s$. So, the neglect of the
dependence of $m_{\eta_{ss}}$ and $m_\phi$ on $m_{u,d}$ is well justified.
In these studies, we performed linear interpolations of
$m_{\pi}^2$ and $m_{\rho}$ at $a=0.20$~fm to determine the values of
$m_{u,d}$ for which $m_\pi/m_\rho$ takes on the values found at $a=0.13$~fm.
Then for $0.13{\rm \ fm}< a < 0.20{\rm \ fm}$ we perform linear
interpolations of the logarithms of $a$, $am_s$, and $am_{u,d}$ in the gauge
coupling along lines of constant $m_\pi/m_\rho$ and $m_{\eta_{ss}}/m_\phi$.

In all figures below we give the values of $m_q/m_s$ for $N_f=3$ and
$m_{u,d}/m_s$ for $N_f=2+1$ at $a=0.13$~fm. The corresponding values of
$m_{\eta_{ss}}/m_\phi$ and $m_\pi/m_\rho$ are given in
Table~\ref{tab:massratio}.

\begin{table}[ht]
\begin{center}
\caption{In the first column we show the value of $m_q/m_s$ at lattice spacing
0.13~fm, which produced the $m_{\eta_{ss}}/m_\phi$ ratio shown in the second
column for spectrum calculations with three equal mass quarks. In the third
column we give the value of $m_{u,d}/m_s$ which produced the $m_\pi/m_\rho$
ratio shown in the fourth column for spectrum calculations with two equal mass
light quarks, and the mass of the heavy quark fixed at $m_s$.
\label{tab:massratio}
}
\vspace{2mm}
\begin{tabular}{|c|c|c|c|}
\hline
\multicolumn{2}{|c|}{\rule[-3mm]{0mm}{8mm}$N_f=3$}&
 \multicolumn{2}{|c|}{\rule[-3mm]{0mm}{8mm}$N_f=2+1$}\\
\hline\hline
\rule[-3mm]{0mm}{8mm}$m_q/m_s$ & $m_{\eta_{ss}}/m_\phi$ &
 $m_{u,d}/m_s$ & $m_\pi/m_\rho$  \\
\hline
\rule[0mm]{0mm}{6mm}1.0 & 0.673   & 1.0  & 0.673    \\
\rule[0mm]{0mm}{0mm}0.6  & 0.583  & 0.6  & 0.582   \\
\rule[0mm]{0mm}{0mm}0.4  & 0.504  & 0.4  & 0.509   \\
\rule[-2mm]{0mm}{2mm}     &        & 0.2  & 0.392   \\
\hline
\end{tabular}
\vspace{-4mm}
\end{center}
\end{table}

\begin{figure}  
\centerline{\includegraphics[width=2.7in]{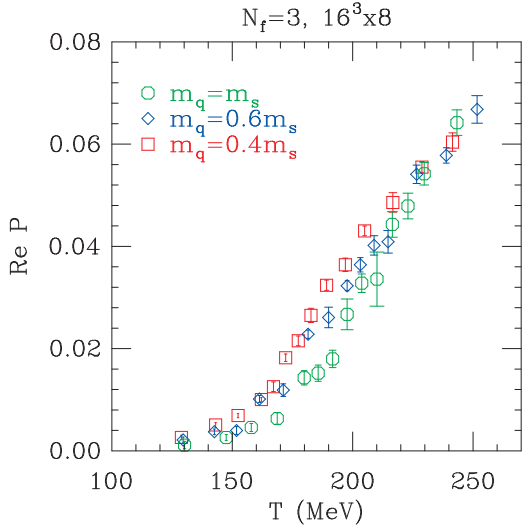}}
\vspace{-7mm}
\caption{Real part of the Polyakov loop on $16^3\times 8$ lattices for
three degenerate flavors of quarks.
\label{fig:rp_nf3_nt8} }
\vspace{-4mm}
\end{figure}

For three equal mass quarks, $N_f=3$, we have carried out thermodynamics
studies on lattices with four, six and eight times slices, and aspect
ratio $N_s/N_t=2$. Here $N_s$ and $N_t$ are the spatial and temporal
dimensions of the lattice in units of the lattice spacing. The spectrum
calculations and interpolations described above allowed us to determine the
values of the quark mass $am_q$ that keep $m_{\eta_{ss}}/m_\phi$ fixed as
the gauge coupling is varied. They also enabled us to determine the value of
the lattice spacing and, therefore, the temperature for each run. 

\begin{figure}
\centerline{\includegraphics[width=2.7in]{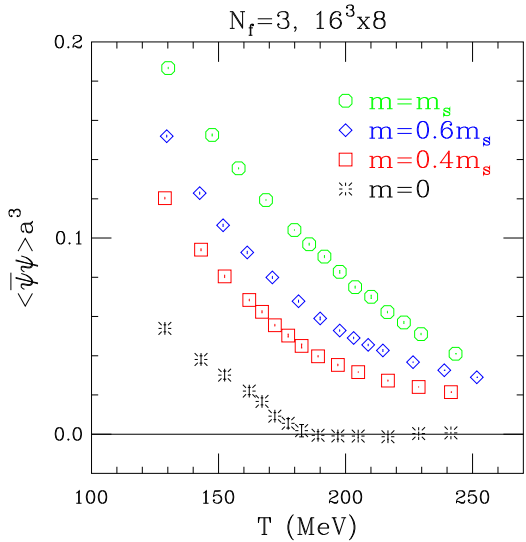}}
\vspace{-7mm}
\caption{The chiral order parameter, $\bar\psi\psi$, 
$16^3\times 8$ lattices.
The black bursts are linear
extrapolations in the quark mass to $m_q=0$ for fixed temperature.
\label{fig:pbp_nf3_nt6_8} }
\vspace{-4mm}
\end{figure}

In Fig.~\ref{fig:rp_nf3_nt8} we plot the real part of the Polyakov loop as
a function of temperature on $16^3\times 8$ lattices. The Polyakov loop shows
a crossover from confined behavior at low temperature to deconfined
behavior at high temperature, but the transition is not particularly sharp.
The insensitivity of the Polyakov loop to the quark mass at fixed temperature
or lattice spacing is perhaps not surprising. We have determined the lattice
spacing from the heavy quark potential, and in our spectrum runs we adjusted
the coupling constant to keep the lattice spacing fixed as the quark mass is
varied. Since the Polyakov loop, like the heavy quark potential, is determined
from measurements of purely gluonic operators, our procedure is likely to
minimize the dependence of the Polyakov loop on the quark mass.

In Fig.~\ref{fig:pbp_nf3_nt6_8} we show the chiral order parameter,
$\bar\psi\psi$ as a function of temperature for $m_q=m_s$, $0.6\,m_s$ and
$0.4\, m_s$ on $16^3\times 8$ lattices. 
The black bursts in 
this figure are linear extrapolations of $\bar\psi\psi$
in the quark mass to $m_q=0$ for fixed temperature. All of these results are
indicative of a crossover at the quark masses studied so far, rather than a
{\it bona fide} phase transition. This result is consistent with earlier studies
which found phase transitions only for smaller values of the quark mass than
we have studied to date \cite{KS_fT}. 
Fig.~\ref{fig:pbp_nf3_nt6_8} suggests 
that there is unlikely to be a phase transition for temperatures
above 180~MeV, but one could occur at or below that value.

The $N_f=2+1$ thermodynamics studies were carried out on $12^3\times 6$
and $16^3\times 8$ lattices. In this phase of our work we performed simulations
with two degenerate light quarks and the heavy quark mass held equal to that of
the strange quark. The spectrum calculations and interpolations described
above allowed us to determine the values of the heavy and light quark masses
needed to keep $m_\pi/m_\rho$ and $m_{\eta_{ss}}/m_\phi$ fixed as the gauge
coupling was varied. They also allowed us to determine the lattice spacing,
and therefore the temperature for each run.

\begin{figure}  
\centerline{\includegraphics[width=2.7in]{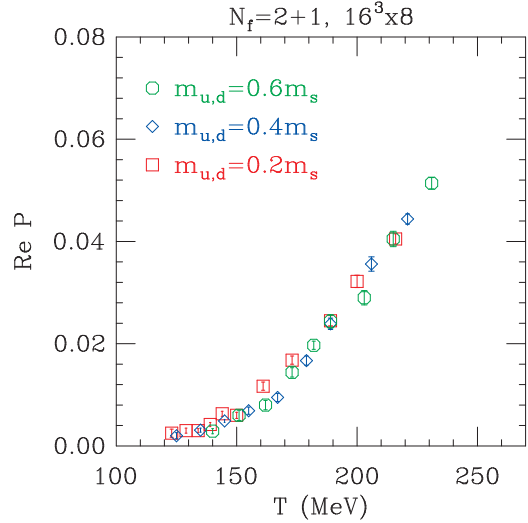}}
\vspace{-7mm}
\caption{Real part of the Polyakov loop on $16^3\times 8$ lattices for
two light and one heavy quark.
\label{fig:rp_nf21_nt8} }
\vspace{-4mm}
\end{figure}

\begin{figure}
\centerline{\includegraphics[width=2.7in]{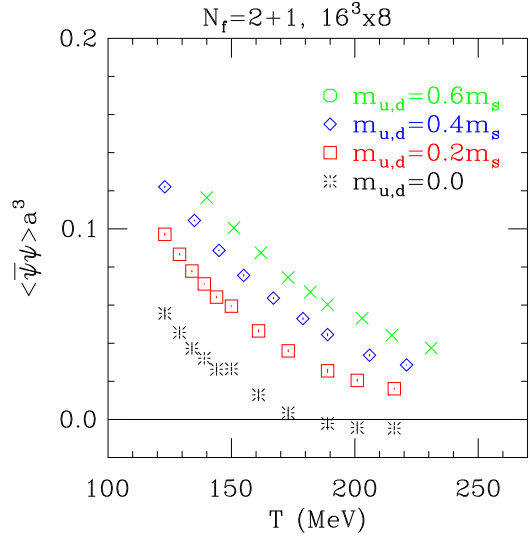}}
\vspace{-7mm}
\caption{The chiral order parameter, $\bar\psi\psi$, on 
$16^3\times 8$ lattices 
The black bursts are linear
extrapolations in the quark mass to $m_q=0$ for fixed temperature.
\label{fig:pbp_nf21_nt6_8} }
\vspace{-4mm}
\end{figure}

In Fig.~\ref{fig:rp_nf21_nt8} we plot the real part of the Polyakov loop
as a function of temperature on $16^3\times 8$ lattices for the three values 
of $m_{u,d}$ studied to date. As in the $N_f=3$, study we observe a crossover
from confined to deconfined behavior, rather than a sharp transition, and
little dependence on the light quark mass.

In Fig.~\ref{fig:pbp_nf21_nt6_8} we plot the chiral order parameter as a
function of temperature on 
$16^3\times 8$ lattices. The
green, blue and red plotting symbols are data for $m_{u,d}=0.6\, m_s$,
$0.4\, m_s$ and $0.2\, m_s$ respectively, and the black bursts are linear
extrapolations in $m_{u,d}$ for fixed temperatures. As in the case of
$N_f=3$, all of these results are indicative of a crossover at the quark
masses studied so far, rather than a {\it bona fide} phase transition.
Fig.~\ref{fig:pbp_nf21_nt6_8} suggests that for
the heavy quark mass equal to that of the strange quark,
there is unlikely to be a phase transtion for $T>170$~MeV.
Simulations at smaller values of $m_{u,d}$ will be
required to determine the precise location and the nature of the expected
finite temperature transition. We plan to carry out such simulations during
the coming year.

\begin{figure}
\centerline{\includegraphics[width=2.7in]{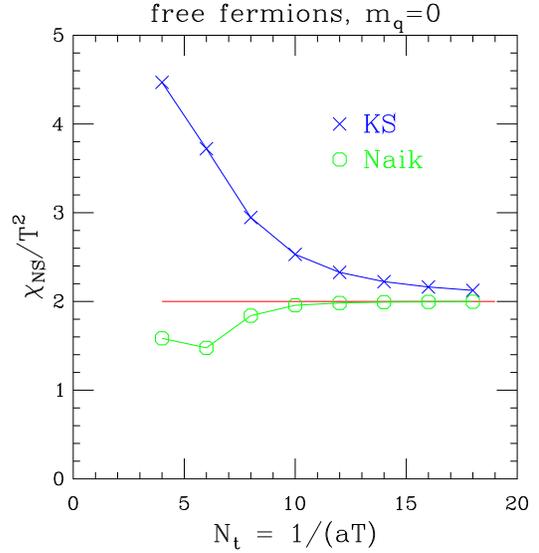}}
\vspace{-7mm}
\caption{The quark number susceptibility for free Asqtad and standard
staggered fermions as function of temporal lattice size $N_t$.
\label{fig:qno_free} }
\vspace{-4mm}
\end{figure}

\begin{figure}
\centerline{\includegraphics[width=2.7in]{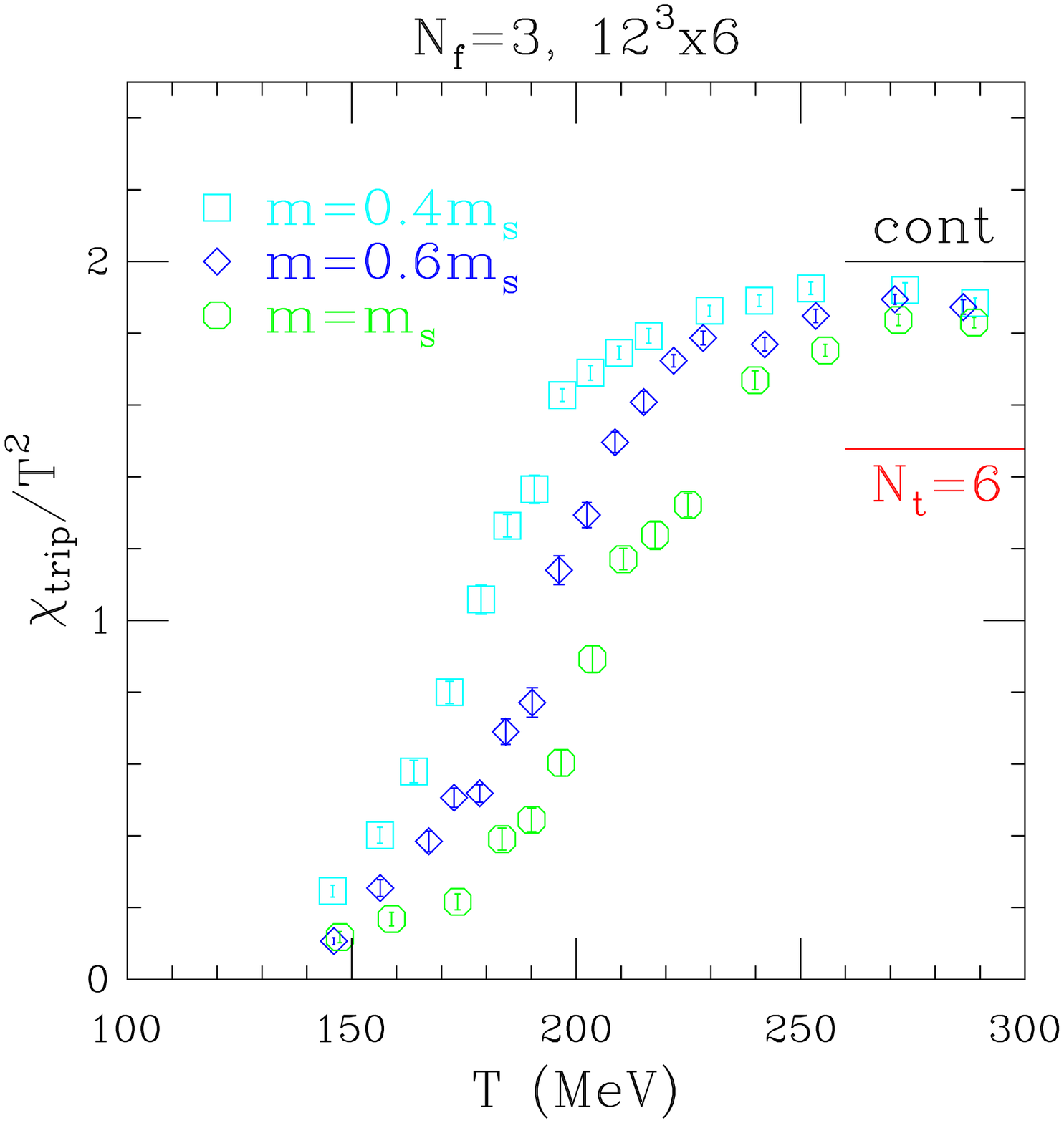}}
\vspace{-7mm}
\caption{The triplet quark number susceptibility for $N_f=3$ on $12^3\times 6$
lattices. The black line on the right of the figure indicates the free quark
value in the continuum, and the red line the free quark value on the
finite lattices used.
\label{fig:qno_nf3_nt} }
\vspace{-4mm}
\end{figure}

\begin{figure}
\centerline{\includegraphics[width=2.7in]{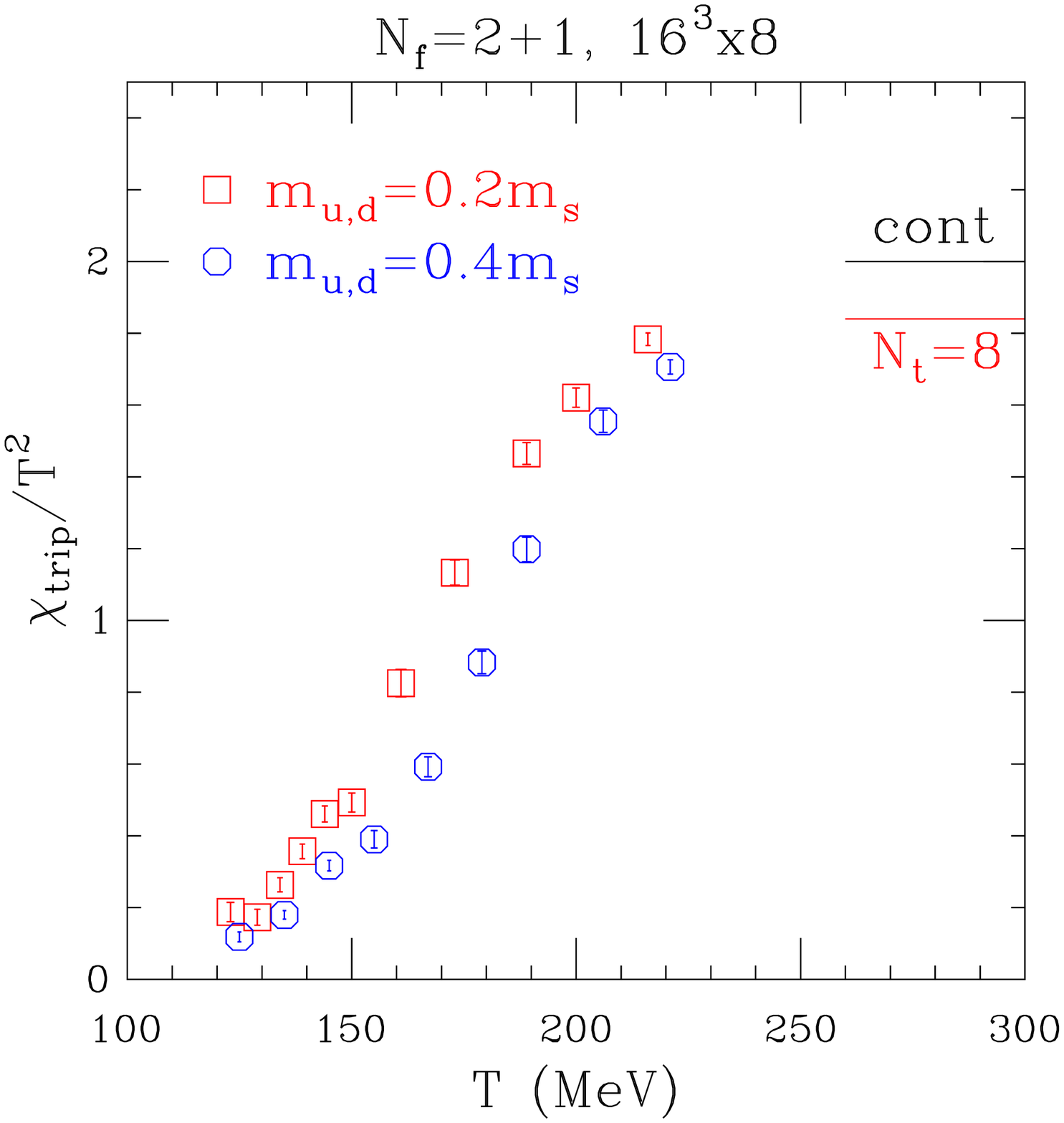}}
\vspace{-7mm}
\caption{The triplet quark number susceptibility for 
$N_f=2+1$ on $16^3\times 8$ lattices. 
The black line on the right of the figure indicates the free quark
value in the continuum, and the red line the free quark value on the
finite lattices used.
\label{fig:qno_nf21_nt} }
\vspace{-4mm}
\end{figure}

We have also measured quark number susceptibilities \cite{SUSC,G_G}.
They can be related to event-by-event fluctuations in heavy ion collisions
\cite{E_by_E} by the fluctuation-dissipation theorem
\be
\left\langle \delta Q^2 \right\rangle \propto \frac{T}{V_s} \frac{\partial^2
 \log Z}{\partial \mu^2_Q} = \chi_Q(T, \mu_Q=0) ~,
\ee
with $\mu_Q$ the chemical potential for the conserved charge $Q$, {\it e.g.}
strangeness. We introduce a chemical potential $\mu$ into the Asqtad action by
\be
&& \!\!\!\!\!\!\!\!\!\! \Dslash_{x,y}(\mu) = \Dslash^{spatial} + \\
&& \!\!\!\!\!\!\!\!\!\!
 \eta_0(x) \left[ U_0^{(F)}(x) {\rm e}^\mu \delta_{x+\hat 0,y} -
 U_0^{(F)\dagger}(x) {\rm e}^{-\mu} \delta_{x,y+\hat 0} \right. \nonumber \\
&& \!\!\!\!\!\!\!\!\!\!
 \left. + U_0^{(L)}(x) {\rm e}^{3\mu} \delta_{x+3\hat 0,y} -
 U_0^{(L)\dagger}(x) {\rm e}^{-3\mu} \delta_{x,y+3\hat 0} \right] ~, \nonumber
\ee
where $U_0^{(F)}$ is the time-like ``fat-link'' and $U_0^{(L)}$ the time-like
``long-link'' including their tadpole improved coefficients.

Defining
\be
\chi_{ij} = \frac{T}{V_s} \frac{\partial^2 \log Z}{\partial \mu_i
 \partial \mu_j} \Bigg|_{\mu=0} ~,
\ee
we computed the light-2-flavor singlet
\be
\chi_{sing} = 2 \chi_{uu} + 2 \chi_{ud}
\ee
and triplet quark number susceptibility
\be
\chi_{trip} = 2 \chi_{uu} - 2 \chi_{ud} ~,
\ee
using stochastic estimates with 100 random Gaussian sources. For $N_f=2+1$
we also computed the strange quark number susceptibility $\chi_{strange}
= \chi_{ss}$.

For free fermions, the Asqtad action, because of the Naik term, reduces
lattice artifacts significantly compared with the standard staggered action
as demonstrated in Fig.~\ref{fig:qno_free}.

\begin{figure}
\centerline{\includegraphics[width=2.7in]{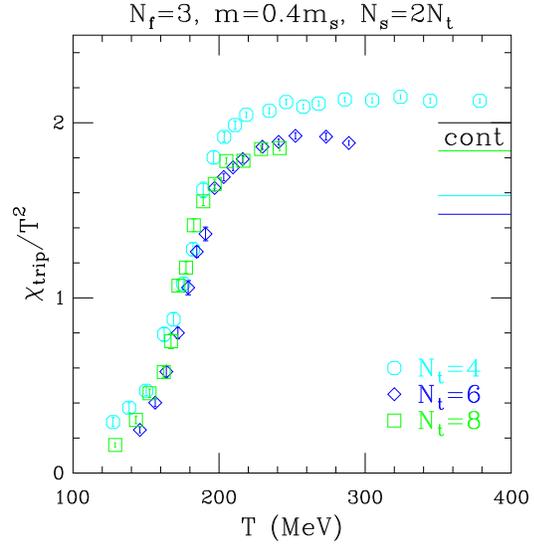}}
\vspace{-7mm}
\caption{The triplet quark number susceptibility for $N_f=3$ with quarks
of mass $m_q=0.4\, m_s$, on $8^3\times 4$, $12^3\times 6$ and $16^3\times 8$
lattices. The black line on the right of the
figure indicates the value for free quarks in the continuum, and the colored
lines the value for free quarks on the finite lattices used.
\label{fig:qno_nf3} }
\vspace{-4mm}
\end{figure}

\begin{figure}
\centerline{\includegraphics[width=2.7in]{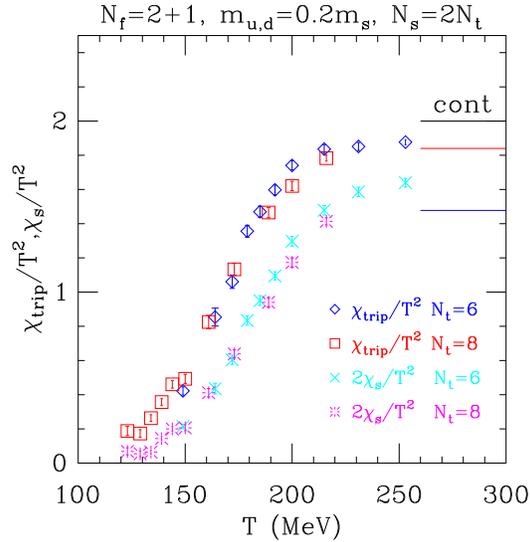}}
\vspace{-7mm}
\caption{The triplet and strange quark number susceptibilities for 
$N_f=2+1$ with $m_{u,d}=0.2\, m_s$ on $12^3\times 6$
and $16^3\times 8$ lattices. The black line on the right of the
figure indicates the value for free quarks in the continuum, and the colored
lines the value for free quarks on the finite lattices used.
\label{fig:qno_nf21} }
\vspace{-4mm}
\end{figure}

In Figs.~\ref{fig:qno_nf3_nt} and \ref{fig:qno_nf21_nt} 
we show the triplet susceptibility for
the three quark masses we have studied in our $N_f=3$ simulations on
$12^3\times 6$ lattices, and for the three values of $m_{u,d}$ we
have studied in our $N_f=2+1$ simulations on $16^3\times 8$ lattices,
respectively.
In Figs.~\ref{fig:qno_nf3} and \ref{fig:qno_nf21} we show the
triplet susceptibility for 
fixed quark mass for the different values of $N_t$ used. 
In Fig.~\ref{fig:qno_nf21} we also show $2\times \chi_{ss}/T^2$,
a quantity which should become equal to $\chi_{trip}/T^2$ in
the very high temperature limit. We observe that
the quark number susceptibility provides a clean signal for the crossover
from the hadronic to the quark-gluon plasma phase. The close agreement
between the $N_t=6$ and 8 results in these figures is an 
indication of the excellent scaling properties in lattice spacing of the
Asqtad action.

\begin{figure}
\centerline{\includegraphics[width=2.7in]{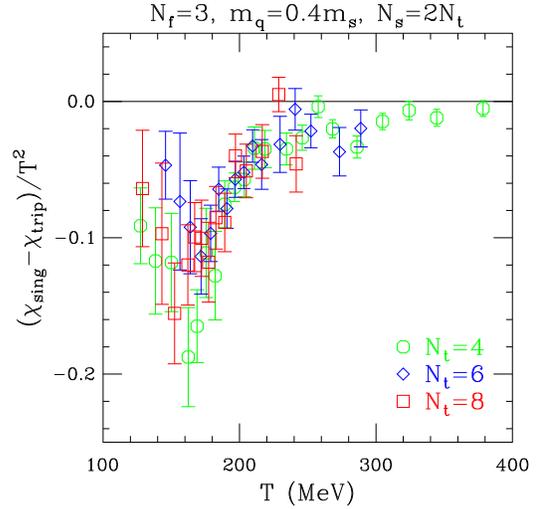}}
\vspace{-7mm}
\caption{The difference between singlet and triplet quark number
susceptibility for $N_f=3$ with quarks of mass $m_q=0.4\, m_s$, on
$8^3\times 4$, $12^3\times 6$ and $16^3\times 8$ lattices.
\label{fig:qnodiff_nf3_m04} }
\vspace{-4mm}
\end{figure}

The singlet susceptibility looks very similar to the triplet susceptibility
shown in Figs.~\ref{fig:qno_nf3} and \ref{fig:qno_nf21}.
We show in Fig.\ref{fig:qnodiff_nf3_m04} the difference between singlet
and triplet susceptibility for $N_f=3$ at the lightest quark mass considered
so far, $m_q=0.4\, m_s$. The two are clearly different in the crossover
region.

This work is supported by the US National Science Foundation and
Department of Energy and used computer resources at Florida State
University (SP), NCSA, NERSC, NPACI, FNAL, and the University of Utah (CHPC).


\begin{thebibliography}{99}

\bibitem{MILC_HT01}
Earlier reports on this work can be found in
C.~Bernard \etal \, (The MILC collaboration),
Nucl.\ Phys.\ A {\bf 702} (2002) 140;
Nucl.\ Phys.\ B (Proc. Suppl) {\bf 106} (2002) 429;

\bibitem{MILC_FATTEST}
K. Orginos, D. Toussaint and R.L. Sugar,
Phys.\ Rev.\ D {\bf 60} (1999) 054503;
Nucl.\ Phys.\ (Proc.\ Suppl.)  {\bf 83} (2000) 878.

\bibitem{LEPAGE98}
G.P. Lepage, Phys.\ Rev.\ D {\bf 59} (1999) 074501.

\bibitem{NAIK}
S. Naik, Nucl. Phys. {\bf B316} (1989) 238.

\bibitem{IMP_SCALING}
C.~Bernard \etal \, (The MILC collaboration),
Phys.\ Rev.\ D {\bf 61} (2000) 111502.

\bibitem{MILC_Spec}
C.~Bernard \etal \, (The MILC collaboration),
Phys.\ Rev.\ D {\bf 64} (2001) 054506.

\bibitem{KS_fT}
S. Aoki \etal \, (JLQCD), Nucl.\ Phys.\ B (Proc. Suppl) {\bf 73} (1999) 459;
F. Karsch, E. Laermann and Ch. Schmidt, Phys.\ Lett.\ {\bf B520} (2001) 41.

\bibitem{SUSC}
S. Gottlieb \etal \, Phys.\ Rev.\ Lett.\ {\bf 59} (1987) 2247.

\bibitem{G_G}
R.V. Gavai and S. Gupta, Phys.\ Rev.\ D {\bf 64} (2001) 074506;
Phys.\ Rev.\ D {\bf 65}, (2002) 094515;
R.V. Gavai, S. Gupta and P. Majumdar, Phys.\ Rev.\ D {\bf 65} (2002) 054506.

\bibitem{E_by_E}
B. M\"uller, Nucl.\ Phys.\ A {\bf 702} (2002) 281;
V. Koch, M. Bleicher and S. Jeon, Nucl.\ Phys.\ A {\bf 702} (2002) 291;
R.V. Gavai, Nucl.\ Phys.\ A {\bf 702} (2002) 299.

\end{thebibliography}
\end{document}